\documentclass[reprint,aps,pre,twocolumn]{revtex4-1}
\usepackage{amsmath,amssymb,mathrsfs}
\usepackage{amsbsy,amsthm,amsfonts}
\usepackage{graphicx}

\allowdisplaybreaks

\begin{document}
\title{Solution of the nonlinear inverse scattering problem by T-matrix completion. I. Theory}

\author{Howard W. Levinson}

\affiliation{Department of Mathematics, University of Pennsylvania, Philadelphia, Pennsylvania 19104, USA}
\email{levh@sas.upenn.edu}

\author{Vadim A. Markel}

\thanks{On leave from the Department of Radiology, University of Pennsylvania, Philadelphia, Pennsylvania 19104, USA}
\email{vmarkel@fresnel.fr,vmarkel@mail.med.upenn.edu}
\affiliation{Aix-Marseille Universit\'{e}, CNRS, Centrale Marseille, Institut Fresnel UMR 7249, 13013 Marseille, France}

\date{\today}

\begin{abstract}
We propose a conceptually new method for solving nonlinear inverse scattering problems (ISPs) such as are commonly encountered in tomographic ultrasound imaging, seismology and other applications. The method is inspired by the theory of nonlocality of physical interactions and utilizes the relevant formalism. We formulate the ISP as a problem whose goal is to determine an unknown interaction potential $V$ from external scattering data. Although we seek a local (diagonally-dominated) $V$ as the solution to the posed problem, we allow $V$ to be nonlocal at the intermediate stages of iterations. This allows us to utilize the one-to-one correspondence between $V$ and the T-matrix of the problem, $T$. Here it is important to realize that not every $T$ corresponds to a diagonal $V$ and we, therefore, relax the usual condition of strict diagonality (locality) of $V$. An iterative algorithm is proposed in which we seek $T$ that is (i) compatible with the measured scattering data and (ii) corresponds to an interaction potential $V$ that is as diagonally-dominated as possible. We refer to this algorithm as to the  data-compatible T-matrix completion (DCTMC). This paper is Part I in a two-part series and contains theory only. Numerical examples of image reconstruction in a strongly nonlinear regime are given in Part II. The method described in this paper is particularly well suited for very large data sets that become increasingly available with the use of modern measurement techniques and instrumentation.
\end{abstract}

\maketitle

\section{Introduction}
\label{sec:intro}

Inverse scattering problems (ISPs) are encountered in optical diffusion tomography~\cite{boas_01_1,arridge_09_1}, diffraction tomography~\cite{bronstein_02_1,devaney_84_1}, electrical impedance tomography~\cite{berryman_90_1,isaacson_04_1,arridge_12_1}, in near-field~\cite{carney_04_1,belkebir_05_1,bao_07_1} and far-field~\cite{chaumet_04_1,belkebir_06_1,mudry_12_1} tomographic electromagnetic imaging, in seismic tomography~\cite{jakobsen_12_1,jakobsen_15_1}, and in many other physical and engineering applications. Solving nonlinear ISPs is a difficult computational task, especially in three dimensions. This is even more true for problems involving large data sets that are available with the use of modern experimental techniques. Developing efficient algorithms for solving nonlinear ISPs remains a fundamental problem of computational physics and an important challenge.

Nonlinear ISPs are amply reviewed in the literature~\cite{newton_book_66,colton_book_98,snieder_98_1,aster_book_05,seo_book_12}. The mainstream approach to solving these problems numerically is Newton's method and its variants such as Levenberg-Marquardt method, iteratively regularized Gauss-Newton method, Newton-Kantorovich method and steepest descent (Landweber iteration). These methods (except for Newton-Kantorovich) are succinctly explained in~\cite{engl_05_1}. Newton-Kantorovich iterations are closely related~\cite{markel_03_2} to the method of inverse Born series~\cite{moskow_08_1,moskow_09_1,arridge_12_1}. A different class of non-deterministic inversion approaches that make use of some form of prior knowledge about the medium is based on Bayesian inference~\cite{watzenig_07_1}. The common feature of all these approaches (except for the inverse Born series) is that a certain cost function is minimized and updated iteratively and that this cost function depends on all available measurements (data points). In the case of inverse Born series, the solution is obtained as an analytically-computable functional of the data.

The method proposed in this paper is conceptually different from the methods reviewed above and is based on a digression into a seemingly unrelated field of physics, namely, into the theory of nonlocality. This theory accounts for the fact that certain physical processes occurring at the point ${\bf r}$ in space can be influenced by the field in some finite vicinity of that point. For example, in local electrodynamics, Ohm's law is written as ${\bf J}({\bf r}) = \sigma({\bf r}) {\bf E}({\bf r})$. In a nonlocal theory, this linear relation is generalized by writing ${\bf J}({\bf r}) = \int V({\bf r}, {\bf r}^\prime) {\bf E}({\bf r}^\prime) d^3 r^\prime$. Of course, we expect on physical grounds that $V({\bf r}, {\bf r}^\prime) \rightarrow 0$ when $\vert {\bf r} - {\bf r}^\prime \vert > \ell$, where $\ell$ is the characteristic scale of nonlocality (the radius of influence), which is usually much smaller than the overall size of the sample. If the electric field ${\bf E}({\bf r})$ does not change noticeably on the scale of $\ell$, we can define the local conductivity as 
\begin{align}
\label{sigma_V}
\sigma({\bf r}) = \int V({\bf r}, {\bf r}^\prime) d^3 r^\prime
\end{align}
\noindent
and use Ohm's law in its local form. This is all well known in physics. However, implications of nonlocality for nonlinear ISPs have not been considered so far.

Let us assume that we want to find $\sigma({\bf r})$ from the measurements of voltage drop for a direct current injected into the sample by two point-like electrodes attached to its surface at various points (Calderon problem). It turns out that it is much easier to find a nonlocal kernel $V({\bf r},{\bf r}^\prime)$ that is consistent with the measurements. Of course, $V({\bf r},{\bf r}^\prime)$ can not be determined uniquely from a typical data set because the number of unknown parameters (degrees of freedom) in $V({\bf r},{\bf r}^\prime)$ is usually much larger than the number of measurements. However, as explained above, we also expect that $V({\bf r}, {\bf r}^\prime)$ should be approximately diagonal. We then proceed as follows:

\begin{enumerate}
  
\item[(1)] First, we define a class of kernels $V({\bf r},{\bf r}^\prime)$ that are compatible with the data. This is the only instance when the data are used, and it turns out that the size of the data set is not a limiting factor for this step.
  
\item[(2)] Then we iteratively reduce the off-diagonal norm of $V({\bf r},{\bf r}^\prime)$ while making sure that $V({\bf r},{\bf r}^\prime)$ remains within the class of ``data-compatible'' kernels.
  
\item[(3)] Once the ratio of the off-diagonal and diagonal norms of $V({\bf r},{\bf r}^\prime)$ is deemed sufficiently small, we compute $\sigma({\bf r}) = \int V({\bf r}, {\bf r}^\prime) d^3 r^\prime$. This gives an approximate numerical solution to the nonlinear ISP. 

\end{enumerate}

\noindent
The above algorithm can be generalized to other ISPs. We refer to it as to Data-Compatible T-matrix Completion (DCTMC) method. 

The role of the T-matrix in solving nonlinear ISPs has been recognized previously~\cite{stolt_80_1,kouri_03_1,jakobsen_12_1,jakobsen_15_1}. The key new insight used in DCTMC is to relax the requirement that $V$ be strictly diagonal. This allows one to establish a one-to-one correspondence between $T$ and $V$. The first advantage of using this approach is that the T-matrix is source- and detector-independent. For example, finite-difference and finite elements forward solvers must be run anew for each source used. The T-matrix approach is free from this requirement. The price of this simplification is that the transformations between $T$ and $V$ involve inversion of dense matrices. However, the computational complexity associated with $T$ to $V$ and $V$ to $T$ operations can be reduced, for example, by exploiting the sparsity of $V$. Second, the T-matrix-based approach results in a useful data reduction, which is applicable to both linear and nonlinear image reconstruction regimes. Finally, the method does not utilize a cost function in the traditional sense and therefore it is not affected by the problem of local minima of the cost function (false solutions).

We underscore that physical interactions are never truly local and some small degree of nonlocality exists in all physical systems. However, the radius of influence $\ell$ is typically so small (e.g., equal to the atomic scale) that the nonlocality can be safely ignored for most practical purposes. In our approach, we relax this condition and allow $V$ to be off-diagonal on much larger scales. Of course, we will seek to find $V$ that is as diagonal as possible. However, we do not expect to eliminate all off-diagonal terms that are separated by more than one atomic scale, not to mention that such fine discretization of the medium is practically impossible. Thus, the non-locality of $V$ that is utilized in DCTMC is not an intrinsic physical property of the material but rather a physically-inspired trick that is used to facilitate the solution of nonlinear ISPs. In other words, we simplify the solution process by relaxing the underlying physical model.

This paper is Part I of a two-part series wherein we focus our attention on theory. Numerical examples for the nonlinear inverse diffraction problem are given in Part II~\cite{PRE_2}. The remainder of this paper is organized as follows. In Sec.~\ref{sec:gen_ISP} we state the general algebraic formulation of the nonlinear ISP that is applicable to many different physical scenarios. In Sec.~\ref{sec:T_matrix}, we introduce the data-compatible T-matrix, which is a central idea of the proposed method. In Sec.~\ref{sec:iter}, we define the basic iterative algorithm of DCTMC. In Sec.~\ref{sec:short}, we introduce ''computational shortcuts'', which combine analytically several steps of the DCTMC algorithm into a single step with a reduced computational complexity. DCTMC algorithm in the linear regime is discussed in Sec.~\ref{sec:lin}. Here we also discuss convergence and regularization of the method. Sec.~\ref{sec:disc} contains a brief discussion. Auxiliary information is given in several appendices. Summary of linearizing approximations (first Born, first Rytov and mean-field) is given in Appendix~\ref{app:lin}. Appendix~\ref{app:M} contains a derivation that establishes the correspondence between DCTMC and the conventional methods in the linear regime. Finally, definitions and properties of several functionals used in this paper are summarized in Appendix~\ref{app:oper}.

\section{General formulation of the ISP}
\label{sec:gen_ISP}

Consider a linear operator ${\mathscr L}$ and the equation
\begin{align}
\label{L_u_S}
{\mathscr L}u({\bf r}) = q({\bf r}) \ , 
\end{align}
\noindent
where $u({\bf r})$ is a physical field and $q({\bf r})$ is the source term. Note that \eqref{L_u_S} does not contain time but can depend parametrically on frequency. It can be said that we work in the frequency domain. Moreover, we consider only a single fixed frequency. Using different working frequencies as additional degrees of freedom for solving an ISP can be very useful (especially if the contrast is approximately frequency-independent, as is often the case in seismic imaging) but is outside of the scope of this paper.

Let ${\mathscr L} = {\mathscr L}_0 - V$, where ${\mathscr L}_0$ is known and $V$ is the unknown interaction operator that we seek to reconstruct. As discussed above, we assume at the outset that $V$ is an integral operator with the kernel $V({\bf r}, {\bf r}^\prime)$ but, eventually, the computed image will be obtained as a function of ${\bf r}$ only. We also assume that $V({\bf r}, {\bf r}^\prime) \neq 0$ only if ${\bf r},{\bf r}^\prime \in \Omega$, where $\Omega$ is a spatial region occupied by the sample. Our goal is to recover $V$ from the measurements of $u$ performed outside of the sample, assuming that it is illuminated by various external sources. We can not perform measurements or insert sources inside the sample, which would have greatly simplified the ISP solution if it was physically possible.

The inverse of ${\mathscr L}$ is the {\em complete} Green's function of the system, denoted by $G = {\mathscr L}^{-1}$. The formal solution to \eqref{L_u_S} is then $u = G q$. We know that $G$ exists as long as the forward problem has a solution. This is usually the case if $V$ is physically admissible. Likewise, the inverse of ${\mathscr L}_0$ is the {\em unperturbed} Green's function, denoted by $G_0 = {\mathscr L}_0^{-1}$. The field $u_{\rm inc} = G_0 q$ is the incident field, in other words, it is the field that would have existed everywhere in space in the case $V=0$. Nonzero $V$ gives rise to a scattered field $u_{\rm scatt}$, and the total field is a sum of the incident and scattered components, $u = u_{\rm inc} + u_{\rm scatt}$. A straightforward algebraic manipulation yields the following result:
\begin{align}
\label{u_s_T_S}
u_{\rm scatt} = (G - G_0) q = G_0 (I - V G_0)^{-1} V G_0 q \ ,
\end{align}
\noindent
where $I$ is the identity operator. 

A single {\em data point} ${\it \Phi}({\bf r}_d, {\bf r}_s)$ is obtained by illuminating the medium with a localized source of unit strength, $q({\bf r}) = \delta({\bf r} - {\bf r}_s)$, and measuring the scattered field by a detector at the location ${\bf r}_d$~\cite{fn1}. By scanning ${\bf r}_d$ and ${\bf r}_s$ on the measurement surfaces $\Sigma_d$ and $\Sigma_s$ outside of the sample, we measure a function of two variables ${\it \Phi}({\bf r}_d, {\bf r}_d)$, which is coupled to $V({\bf r}, {\bf r}^\prime)$ by the equation
\begin{align}
\label{Phi_V}
G_0 (I - V G_0)^{-1} V G_0 = {\it \Phi} \ .
\end{align}
\noindent
All product and inversion operations in \eqref{Phi_V} should be understood in the operator sense. The ISP can now be formulated as follows: Given a measured function ${\it \Phi}({\bf r}_d, {\bf r}_s)$, where ${\bf r}_d \in \Sigma_d$ and ${\bf r}_s \in \Sigma_s$, find an ``approximately diagonal'' kernel $V({\bf r}, {\bf r}^\prime)$, where ${\bf r}, {\bf r}^\prime \in \Omega$. We do not need to define ``approximate diagonality'' precisely at this point, but in the case of matrices that are inevitably used in all computations, this requirement implies a sufficiently small ratio of the off-diagonal and diagonal norms.

\begin{figure*}
\centerline{\includegraphics[width=16.4cm]{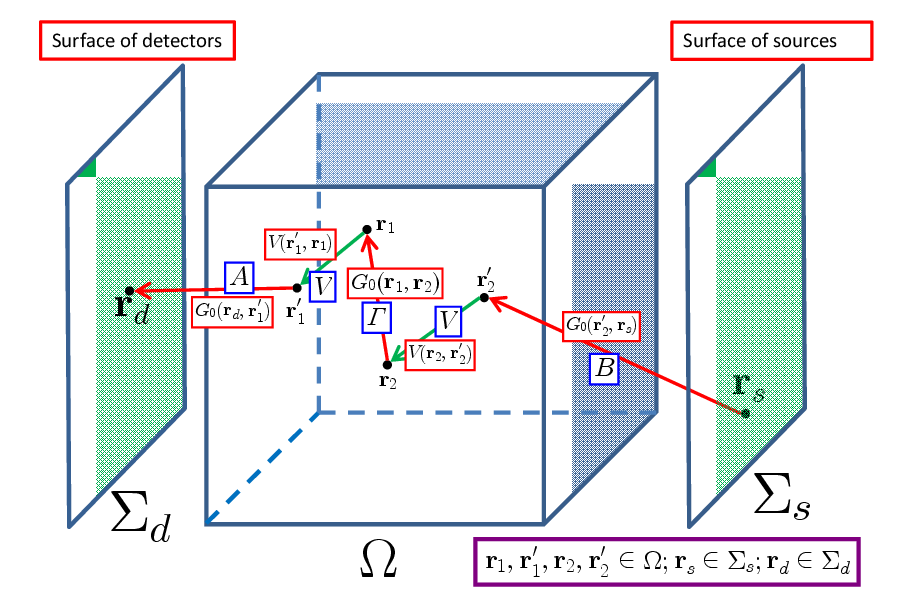}}
\caption{\label{fig:geom} (color online)
Illustration of the imaging geometry. The symbols $A$, $B$, ${\it \Gamma}$\ and $V$ in the rectangular frames denote the matrices obtained by restricting and sampling the kernels $G_0({\bf r}, {\bf r}^\prime)$ and $V({\bf r}, {\bf r}^\prime)$. The scattering diagram corresponds to the second order term $G_0VG_0VG_0$ in the formal power-series expansion of the left-hand side in \eqref{Phi_V}. Note that in the local limit $V({\bf r},{\bf r}^\prime) = D({\bf r})\delta({\bf r} - {\bf r}^\prime)$, the two arrows contract to two vertexes at ${\bf r}_1 = {\bf r}_1^\prime$ and ${\bf r}_2 = {\bf r}_2^\prime$.}
\end{figure*}

It is important to note that $G_0$ in \eqref{Phi_V} is the same operator in all instances where it appears, but for the purpose of computing the operator products and inverses, its kernel $G_0({\bf r}, {\bf r}^\prime)$ is differently restricted. This is illustrated graphically in Fig.~\ref{fig:geom}. Thus, for the first term $G_0$ in the left-hand side of \eqref{Phi_V}, ${\bf r}={\bf r}_d \in \Sigma_d$ and ${\bf r}^\prime ={\bf r}_1^\prime \in \Omega$.  For the second term (inside the brackets) ${\bf r} = {\bf r}_1 \in \Omega$ and ${\bf r}^\prime = {\bf r}_2 \in \Omega$. For the last term, ${\bf r} = {\bf r}_2^\prime \in \Omega$ and ${\bf r}^\prime = {\bf r}_s \in \Sigma_s$. We emphasize that the imaging geometry shown in Fig.~\ref{fig:geom} is representative but not very general. In particular, the measurement surfaces $\Sigma_d$ and $\Sigma_s$ can be larger or smaller than the face of the cube, or curve, or even be regions of space of finite volume rather than surfaces~\cite{fn2}. The sample volume $\Omega$ does not have to be cubic and, in an extreme case, it can be a two-dimensional surface. All this has no bearing on the method of this paper. The only requirement that we impose, which is physical rather than mathematical, is that $\Sigma_d$ and $\Sigma_s$ do not overlap with $\Omega$. However, $\Sigma_d$ can overlap with $\Sigma_s$.

Further, in all practical implementations, the data are sampled rather than measured continuously and the medium is voxelized. An example of such discretization is given in~\cite{PRE_2}. At this point we proceed under the assumption that \eqref{Phi_V} can be suitably discretized and converted to a matrix equation. In this case, it is logical to use different notations for the matrices that are obtained by different restriction and sampling of the kernel $G_0({\bf r}, {\bf r}^\prime)$.  Indeed, the matrices obtained in this manner are different and can even be of different size. We will denote the matrices obtained by sampling the first, the second, and the last terms $G_0$ in \eqref{Phi_V} by $A$, ${\it \Gamma}$ and $B$, respectively. These notations are also illustrated in Fig.~\ref{fig:geom}. Then the discretized version of \eqref{Phi_V} takes the following form:
\begin{align}
\label{Phi_V_matrix}
A (I - V{\it \Gamma})^{-1} V B = {\it \Phi} \ .
\end{align}
In \eqref{Phi_V_matrix}, $A$, $B$ and ${\it \Gamma}$ are known theoretically, ${\it \Phi}$ is measured, and we seek to find the unknown $V$.

Eq.~\eqref{Phi_V_matrix} is the main nonlinear equation that is discussed in this paper. It is, in fact, very general and encompasses
many different problems of imaging and tomography. The underlying physical model is encoded in the operator $G_0$ and in the matrices
$A$, $B$ and ${\it \Gamma}$ that are obtained by sampling this operator. The following three important remarks about this equation can be made:\\

{\bf Remark 1: Noninvertibility of $A$ and $B$.} If matrices $A$ and $B$ were invertible in the ordinary sense, the nonlinear ISP would be solvable exactly by three operations of matrix inversion.  Unfortunately, $A$ and $B$ are almost never invertible. To construct $A$ and $B$ of sufficiently high rank, one needs to perform measurements inside the medium. As was noted above, this is usually impossible. The typical sizes of all matrices involved will be discussed below (see Fig.~\ref{fig:matrix} and its discussion). \\

{\bf Remark 2: Linearization.} One may seek a linearization of the ISP by approximating the left-hand side in \eqref{Phi_V_matrix} with various expressions that allow an analytical linearizing transformation. The three main approaches to achieve this end are first Born, first Rytov and mean-field approximations, and they are briefly summarized in Appendix~\ref{app:lin}. In the mathematical formulation of the ISP, the three approximations differ only by the transformation that is applied to the data matrix ${\it \Phi}$, while the general form of the linearized equation is in all cases
\begin{align}
\label{Psi_V_matrix_lin}
A V B  = {\it \Psi}[{\it \Phi}] \ ,
\end{align}
\noindent
where ${\it \Psi}[{\it \Phi}]$ is the appropriate transformation of the data matrix; in the simplest case of first Born approximation, ${\it \Psi}[{\it \Phi}] = {\it \Phi}$.\\

{\bf Remark 3: Matrix unrolling for the linearized problem.} The linearization approaches described in Appendix~\ref{app:lin} do not require or enforce by design the  diagonality of $V$. However, in the conventional treatments of the problem, it is typical to assume that $V$ is strictly diagonal and to operate with the vector $\vert \upsilon \rangle$ composed of the diagonal elements of $V$. Accordingly, the matrix ${\it \Psi}$ is unrolled into a vector $\vert \psi \rangle$ by the matrix operation known as {\tt vec}, that is, by stacking the columns of ${\it \Psi}$ into one column-vector. The resultant equation has the form 
\begin{align}
\label{K_phi_lin}
K \vert \upsilon \rangle = \vert \psi \rangle \ , 
\end{align}
\noindent
where $K$ is a matrix obtained by multiplying the elements of $A$ and $B$ according to the rule $K_{(mn),j} = A_{mj}B_{jn}$ and $(mn)$ is a composite index. The important point here is that the conventional methods often treat $K$ in \eqref{K_phi_lin} as a matrix of the most general form. In contrast, DCTMC algorithm takes account of the special algebraic structure of $K$ and, therefore, can be used advantageously even in the linear regime. This is discussed in more detail in Appendix~\ref{app:M}.

\section{T-matrix and its representations; ``experimental'' T-matrix}
\label{sec:T_matrix}

The basic definition of the T-matrix (which is, actually, an operator) is through the relation between the complete and the unperturbed Green's functions: $G = G_0 + G_0 T G_0$. By direct comparison with \eqref{u_s_T_S} we find that
\begin{align}
\label{T_def}
T = (I - V G_0)^{-1} V \ .
\end{align}
\noindent
We will not use different notations for the operator $T$ and its discretized version, which is truly a matrix. Consequently, Eq.~\eqref{Phi_V_matrix} can be rewritten as
\begin{align}
\label{Phi_T}
 A T B = {\it \Phi} \ ,
\end{align}
\noindent
where
\begin{align}
\label{T_from_V}
T = {\cal T}[V] \equiv (I - V {\it \Gamma})^{-1} V = V(I - {\it \Gamma}V)^{-1} \ .
\end{align}
\noindent
Here we have defined the nonlinear functional ${\cal T}[\cdot]$, which contains ${\it \Gamma}$ as a parameter. 

We can view \eqref{T_from_V} as a matrix formulation of the forward problem. If $V$ is known, we can use \eqref{T_from_V} to compute $T$, and once this is accomplished, we can predict the result of a measurement by any detector due to any source by matrix multiplication according to \eqref{Phi_T}. Therefore, computation of $T$ yields the most general solution to the forward problem. The forward solution is usually known to exist if $V$ is physically admissible. In the iterative process of DCTMC, we can ensure physical admissibility of $V$ every time before the transformation ${\cal T}[V]$ is used. We can view this procedure as a particular type of regularization by imposition of physical constraints. If this type of regularization is used, one can be sure that the matrix inversion involved in computing ${\cal T}[V]$ is always well-defined.

We can also formally invert ${\cal T}$ and write 
\begin{align}
\label{V_from_T}
V = {\cal T}^{-1}[T] \equiv (I + T {\it \Gamma})^{-1} T = T(I + {\it \Gamma}T)^{-1} \ .
\end{align}
\noindent
Much less is known about the existence of the inverse in \eqref{V_from_T}. In other words, we do not know the conditions of physical admissibility of $T$ apart from the general but not very useful symmetry property $T_{ij} = T_{ji}$. Certainly, ${\cal T}^{-1}[T]$ does not exist for all arguments $T$. In DCTMC, one of the possible approaches is to update $V$ iteratively by using \eqref{V_from_T}. In this case, existence of the inverse is required. While we do not possess a general proof, numerical simulations for the inverse diffraction problem have encountered no singularities in \eqref{V_from_T}. More importantly, the problem of invertibility of ${\cal T}$ does not arise at all if Computational Shortcut 2 is used (see Sec.~\ref{sec:short.T2D} below).

\begin{figure}
\centerline{\includegraphics[width=8.2cm]{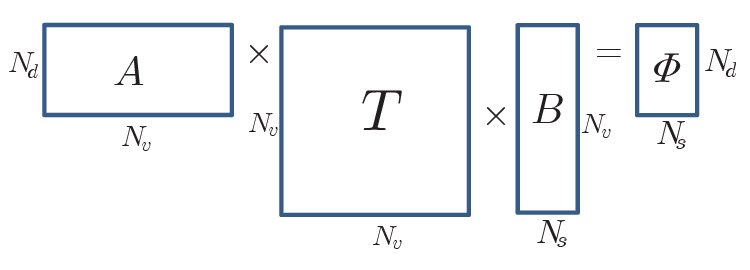}}
\caption{\label{fig:matrix} (color online) Block diagram of Eq.~\eqref{Phi_T} with sizes of all matrices indicated. Here $N_d$ and $N_s$ are the numbers of detectors and sources and $N_v$ is the number of voxels.}
\end{figure}

A block diagram of Eq.~\eqref{Phi_T} with all matrix sizes indicated is shown in Fig.~\ref{fig:matrix}. Here $N_d$ and $N_s$ are the numbers of detectors and sources used (not necessarily equal) and $N_v$ is the number of volume voxels. For a practical estimate of these numbers, refer to Fig.~\ref{fig:geom}. Let the measurement surfaces $\Sigma_d$ and $\Sigma_s$ be identical squares located on the opposite sides of a cubic sample. Let the detectors and sources be scanned on an $L \times L$ square grids and let the sample be discretized on a $L \times L \times L$ cubic grid with the same pitch. Then $N_d = N_s = L^2$, $N_v=L^3$. These estimates are typical but, admittedly, not very general. Still, in many practical cases we can expect that
\begin{align}
\label{estimates}
N_d,N_s \ll N_v \ll N_d N_s \ .
\end{align}
\noindent
The first inequality in \eqref{estimates} illustrates Remark 1 of Sec.~\ref{sec:gen_ISP} because the matrices $A$ and $B$ are in this case clearly not invertible. The second inequality is important if we wish to compare DCTMC to some of the traditional approaches. For example, the conventional formulation of the linearized ISP starts from equation \eqref{K_phi_lin}. As is explained in Remark 3, is is commonly assumed that $K$ in \eqref{K_phi_lin} is a general matrix of the size $N_d N_s \times N_v$ ($L^4 \times L^3$). However, the sizes of $A$ and $B$ are $N_d \times N_v$ ($L^2\times L^3$) and $N_v \times N_s$ ($L^3 \times L^2$), respectively. Computing numerically the pseudoinverse of $K$ (if we do not account for its special algebraic structure as is described in Appendix~\ref{app:M}) is a much more computationally-intensive task than computing the pseudoinverses of $A$ and $B$. Therefore, the relaxation of the strict requirement of diagonality of $V$ allows one to work with two much smaller ``weight matrices'' $A$ and $B$ instead of one large ``weight matrix'' $K$. 

We now turn to the central idea of DCTMC, namely, to the concept of data-compatibility of the T-matrix. To formulate the constraints that equation \eqref{Phi_T} places on $T$ in a computationally-tractable form, consider the singular value decompositions of $A$ and $B$:
\begin{align}
\label{SVD}
A = \sum_{\mu=1}^{N_d} \sigma^A_\mu \left\vert f^A_\mu \right\rangle \left\langle g^A_\mu \right\vert \ , \ \ B = \sum_{\mu=1}^{N_s} \sigma^B_\mu \left\vert f^B_\mu \right\rangle  \left\langle g^B_\mu \right\vert \ . 
\end{align}
\noindent
Here $\sigma_\mu^A$, $\left\vert f_\mu^A \right\rangle$ and $\left\vert g_\mu^A \right\rangle$ are the singular values and right and left singular vectors of $A$, and similarly for $B$. Note that $\left\vert f_\mu^A \right\rangle$ and $\left\vert g_\mu^B \right\rangle$ are vectors of length $N_d$ and $N_s$, respectively, while $\left\vert g_\mu^A \right\rangle$ and $\left\vert f_\mu^B \right\rangle$ are both of length $N_v$, and we have assumed in \eqref{SVD} that $N_d,N_s \leq N_v$. Using the orthogonality of singular vectors, we obtain from \eqref{Phi_T} and \eqref{SVD}
\begin{align}
\label{Psi_Phi}
\sigma_\mu^A \sigma_\nu^B \tilde{T}_{\mu\nu} = {\it \tilde{\Phi}}_{\mu\nu} \ , \ \ 1\leq \mu \leq N_d \ , \ 1\leq \nu \leq N_s \ ,
\end{align}
\noindent
where
\begin{subequations}
\label{Psi_Phi_defs}
\begin{align}
\label{T_tilde_def}
& \tilde{T}_{\mu\nu} \equiv \left\langle g_\mu^A \vert T  \vert f_\nu^B \right \rangle \ , \ \ 1\leq \mu,\nu \leq N_v \ ; \\ 
\label{Phi_tilde_def}
& {\it \tilde{\Phi}}_{\mu\nu} \equiv \left\langle f_\mu^{A} \vert {\it \Phi} \vert g_\nu^B \right\rangle \ , \ 1\leq \mu \leq N_d \ , \ 1 \leq \nu \leq N_s \ .
\end{align}
\end{subequations}
\noindent
By $\tilde{T}$ we denote the T-matrix in {\em singular-vector representation} while $T$ that was used previously is the T-matrix in {\em real-space representation}. The two representations are related to each other by the transformation
\begin{align}
\label{Rotation}
\tilde{T} = R_A^{*} T R_B \equiv {\cal R}[T]\ , \ \ T = R_A \tilde{T} R_B^{*} \equiv {\cal R}^{-1}[\tilde{T}]\ , 
\end{align}
\noindent
where $R_A$ is the unitary matrix whose columns are the singular vectors $\left\vert g_\mu^A \right\rangle$ while $R_B$ is the unitary matrix whose columns are the singular vectors $\left\vert f_\mu^B \right\rangle$. Equation \eqref{Rotation} defines the pseudo-rotation functional ${\cal R}[\cdot]$. We note that ${\cal R}[\cdot]$ is linear and invertible even though it is not equivalent to a conventional rotation because $R_A \neq R_B$. It is useful to keep in mind that ${\it \tilde{\Phi}} \neq {\cal R}[{\it \Phi}]$. As can be seen from the definition \eqref{Psi_Phi_defs}, ${\it \tilde{\Phi}}$ is related to ${\it \Phi}$ by a similar transformation but with different unitary matrices. 

We now can write the solution to \eqref{Psi_Phi} as follows:
\begin{align}
\label{T_inv}
\tilde{T}_{\mu\nu} = \left\{
\begin{array}{ll}
{\displaystyle \frac{1}{\sigma_\mu^A \sigma_\nu^B} {\it \tilde{\Phi}}_{\mu \nu}} 
& \ , \ \ {\rm if} \ \sigma_\mu^A \sigma_\nu^B >    \epsilon^2 \ ; \\
{\rm unknown} 
& \ , \ \ {\rm otherwise} \ \ .
\end{array} \right. 
\end{align}
\noindent
Here $\epsilon$ is a small positive constant. If computations could be performed with infinite precision, we could have set $\epsilon=0$. In practice, we should take $\epsilon$ to be small but at least larger than the smallest positive floating-point constant for which a particular implementation of numerical arithmetic adheres to the IEEE standard. We note that under the assumptions stated above, the condition $\sigma_\mu^A \sigma_\nu^B > \epsilon^2$ can be satisfied only for $1\leq \mu \leq N_d$ and $1 \leq \nu \leq N_s$. Singular values $\sigma_\mu^A$ and $\sigma_\nu^B$ with indexes outside of these ranges are identically zero.

\begin{figure}
\centerline{\includegraphics[width=8.2cm]{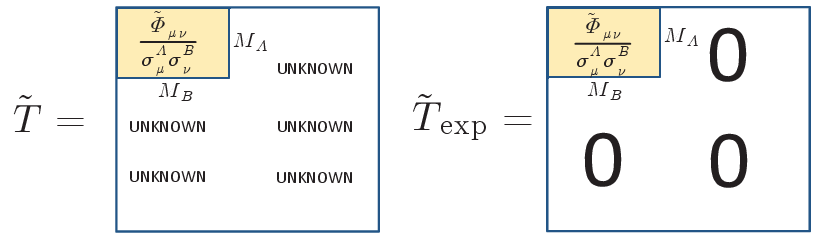}}
\caption{\label{fig:T} (color online) Left panel: Elements of $\tilde{T}$ inside the shaded block can be computed from the data by using \eqref{T_inv}. Elements outside of the shaded block are not known and can not be in any way inferred from the data. Right panel: The initial guess for the T-matrix, $T_{\rm exp}$. In this initial guess, we set the unknown elements of $\tilde{T}$ (in singular-vector representation) to zero.}
\end{figure}

\begin{figure}
\centerline{\includegraphics[width=8.2cm]{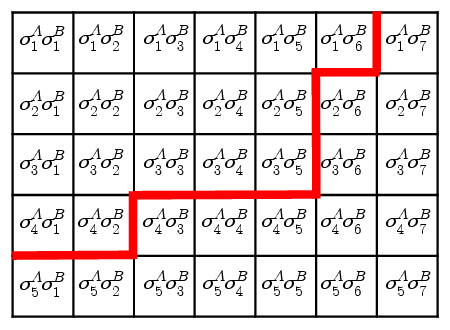}}
\caption{\label{fig:reg} (color online) Assuming that the singular values $\sigma_\mu^A$ and $\sigma_\mu^B$ are arranged in the descending order, this sketch shows an example of a more general shape (compared to Fig.~\ref{fig:T}) of the region in which the elements of $\tilde{T}$ are known. The numbers above the thick line satisfy $\sigma_\mu^A \sigma_\nu^B > \epsilon^2$. In the general case, the boundary line can only go from left to right and from bottom to top if followed from the left-most boundary of the matrix.}
\end{figure}

Eq.~\eqref{T_inv} summarizes our knowledge about the system that is contained in the data. There are few matrix elements of $\tilde{T}$ that are known with certainty. These matrix elements can be computed by the first expression in~\eqref{T_inv}. The other matrix elements can not be determined from equation \eqref{Phi_T}. We can vary these unknown elements arbitrarily and the error of \eqref{Phi_T} will not noticeably change. The number of known elements of $\tilde{T}$ can not exceed $N_d N_s$ but can, in principle, be smaller, e.g., if the rank of $A$ is less than $N_d$ or the rank of $B$ is less than $N_s$, although this situation is not typical even for severely ill-posed ISPs. In any event, $N_d N_s$ is usually much smaller than the total number of the matrix elements of $\tilde{T}$, which is equal to $N_v^2$. Using the previously introduced estimates, $N_d N_s / N_v^2 \sim 1/L^2$. Therefore, only a small fraction of the elements of $\tilde{T}$ are known. In what follows, we assume that the singular values of $A$ and $B$ are arranged in the descending order and that the known elements of $\tilde{T}$ can be collected into the upper-left rectangular block of the size $M_A \times M_B$ (see Fig.~\ref{fig:T}), where $M_A \leq N_d$ and $M_B\leq N_s$. We emphasize again that, in many practically-important cases, equalities will hold in the above expressions. However, it is possible to arrange the sources in such a way that the rank of $B$ is less than $N_s$ (and similarly for detectors and $A$), at least up to the numerical precision of the computer~\cite{fn3}. Moreover, the region of known matrix elements can be of a more general shape than a rectangle, as is shown in Fig.~\ref{fig:reg}. It is not conceptually difficult to account for this fact. However, we will proceed under the assumption that the region is rectangular in order to shorten the discussion. Besides, in the numerical simulations of~\cite{PRE_2}, this region was, in fact, rectangular. 

Even though we can not gain any knowledge about the matrix elements of $\tilde{T}$ outside of the shaded area shown in Fig.~\ref{fig:T} by using equation \eqref{Phi_T} alone, we can make an initial guess for $\tilde{T}$, which we denote by $\tilde{T}_{\rm exp}$ (the ''experimental'' T-matrix). We define $T_{\rm exp}$ (in real-space representation) as the matrix that satisfies \eqref{Phi_T} in the minimum norm sense and has the smallest entry-wise norm $\Vert T \Vert_2$. This matrix is uniquely defined by the equation
\begin{align}
	\label{T_exp_Phi}
	T_{\rm exp} = A^+ {\it \Phi} B^+ \ ,
\end{align}
\noindent
where ``$+$'' denotes Moore-Penrose pseudoinverse. If $A$ and $B$ are rank-deficient or invertible, \eqref{Phi_T} is satisfied by $T_{\rm exp}$ exactly so that $\Vert A T_{\rm exp} B - {\it \Phi} \Vert_2 = 0$. The experimental T-matrix in singular-vector representation is obtained from \eqref{T_exp_Phi} by the transformation \eqref{Rotation}. In fact, the experimental T-matrix is more easily characterizable in singular-vector representation. Indeed, all the elements of $\tilde{T}_{\exp}$ in the unshaded area of the diagram in Fig.~\ref{fig:T} (right panel) are equal to zero. This is expressed mathematically by writing 
\begin{align}
\label{T_exp}
\left(\tilde{T}_{\rm exp} \right)_{\mu\nu} = \left\{
\begin{array}{ll}
{\displaystyle \frac{1}{\sigma_\mu^A \sigma_\nu^B} {\it \tilde{\Phi}}_{\mu \nu}} 
& \ , \ \ {\rm if} \ \sigma_\mu^A \sigma_\nu^B >    \epsilon^2 \ ; \\
0 
& \ , \ \ {\rm otherwise} \ \ .
\end{array} \right.
\end{align}
\noindent
This expression is equivalent to \eqref{T_exp_Phi}. 

We conclude this section with two important observations about the experimental T-matrix:\\

{\bf Remark 4: Lack of sparsity of $T_{\rm exp}$.} The matrix $\tilde{T}_{\rm exp}$ is sparse but the same is not true for $T_{\rm exp}$.\\

{\bf Remark 5: Lack of symmetry of $\tilde{T}_{\rm exp}$.} It is known theoretically that the correct T-matrix is symmetric in real-space representation. However, this is not generally true for $T_{\rm exp}$. Indeed, $T_{\rm exp} = {\cal R}^{-1}[\tilde{T}_{\rm exp}]$, and in $\tilde{T}_{\rm exp}$ a large fraction of the elements are replaced by zeros. The resultant $T_{\rm exp}$ is not likely to be symmetric.\\

\section{Basic iteration cycle}
\label{sec:iter}

In this section we describe a computational algorithm in which the matrices $T$ and $V$ are continuously updated so that $T$ is kept data-compatible and $V$ becomes increasingly diagonally-dominated. Our goal is to fill the unknown elements of $\tilde{T}$ (the white areas in the left panel of Fig.~\ref{fig:T}) in such a way that the corresponding interaction matrix $V$, computed according to \eqref{V_from_T}, is approximately diagonal. This is a general formulation of the problem of matrix completion, although the constraint that we apply to $\tilde{T}$ is not the same as in the conventional statement of the problem. 

Before proceeding, we need to introduce several additional operators. First, define the masking operators ${\cal M}[\cdot]$ and ${\cal N}[\cdot]$:
\begin{subequations}
\begin{align}
\label{Mask}
\left({\cal M}[\tilde{T}] \right)_{\mu\nu} \equiv
 \left\{
 \begin{array}{ll}
 0\ ,   & \sigma_\mu^A \sigma_\nu^B > \epsilon^2  \ \ ; \\
 \tilde{T}_{\mu\nu}\ , & {\rm otherwise} \ .
 \end{array} \right. \\
 \left({\cal N}[\tilde{T}] \right)_{\mu\nu} \equiv
 \left\{
 \begin{array}{ll}
 \tilde{T}_{\mu\nu} \ ,   & \sigma_\mu^A \sigma_\nu^B > \epsilon^2  \ \ ; \\
 0 \ , & {\rm otherwise} \ .
\end{array} \right.
\end{align}
\end{subequations}
\noindent
We note that ${\cal M}[\tilde{T}] + {\cal N}[\tilde{T}] = \tilde{T}$. Then the operator of enforcing data-compatibility of $\tilde{T}$ (in singular-vector representation) ${\cal O}[\cdot]$ can be defined as follows:
\begin{align}
\label{O_def}
{\cal O}[\tilde{T}] \equiv {\cal M}[\tilde{T}] + \tilde{T}_{\rm exp} = \tilde{T} -  {\cal N}[\tilde{T}] + \tilde{T}_{\rm exp} \ .
\end{align}
\noindent
It can be seen that the action of ${\cal O}[\tilde{T}]$ is to overwrite (hence the notation ${\cal O}$) the elements of $\tilde{T}$ in the shaded area of Fig.~\ref{fig:T} with the elements of $\tilde{T}_{\rm exp}$ and to leave all other elements unchanged. The operator ${\cal O}[\cdot]$ is defined for any $N_v \times N_v$ matrix but in the iterations discussed below we always apply this operator to the T-matrix in singular-vector representation. 

Next, we will need to define a diagonal approximation to $V$. To this end, we define an entry-wise ``force-diagonalization'' operator ${\cal D}[\cdot]$, where
\begin{align}
\label{D_def}
\left({\cal D}[V] \right)_{ij} \equiv
\delta_{ij} \sum_k  V_{ik} \rho(\ell_{ki}) \ .
\end{align}
\noindent
Here $\rho(\ell_{ki})$ is a weight function that depends on the physical distance $\ell_{ik}$ between the voxels $i$ and $k$ and not on the relative position of the lines $i$ and $k$ in the matrix $V$. The definition \eqref{D_def} is in agreement with the approximation of the form \eqref{sigma_V}. We note that a more symmetric definition involving the factor $\rho(\ell_{ki}) (V_{ik}  + V_{ki}) / 2$ seems to be more natural to use but, in fact, this expression is not of the same form as \eqref{sigma_V} and we have verified numerically that it does not produce superior results compared to \eqref{D_def}. If $V$ is symmetric, the two expressions yield identical results but the iteratively-updated $V$ is not symmetric in DCTMC, and the physical meaning of the off-diagonal elements $V_{ik}$ and $V_{ki}$ is generally not the same. In the simplest case, we can take $\rho(\ell_{ki}) = \delta_{ki}$. This corresponds to sending all off-diagonal elements of $V$ to zero. This approach allows for a complete and simple analysis of DCTMC in the linear regime, as is described in Sec.~\ref{sec:lin} below. However, the use of more complicated functions $\rho(\ell)$ corresponds better to the spirit of DCTMC; it allows one to take the off-diagonal elements of $V$ that are generated in the course of iterations. Unlike the operator ${\cal O}[\cdot]$, ${\cal D}[\cdot]$ will always be applied in real-space representation. 

We are now ready to describe the iterative process of DCTMC. The iteration steps will be defined in terms of the operators ${\cal R}[\cdot]$ and ${\cal T}[\cdot]$, ${\cal D}[\cdot]$ and ${\cal O}[\cdot]$. A summary of all operators used in this paper is given in Appendix~\ref{app:oper}. We assume that the SVD decompositions of matrices $A$ and $B$ and the experimental T-matrix $\tilde{T}_{\rm exp}$ \eqref{T_exp} have been precomputed. Consider the case when the iterations start from an initial guess for the T-matrix. We then set $k=1$, $\tilde{T}_1 = \tilde{T}_{\rm exp}$ and run the following iterations:

\begin{figure*}
\centerline{\includegraphics[width=16.4cm]{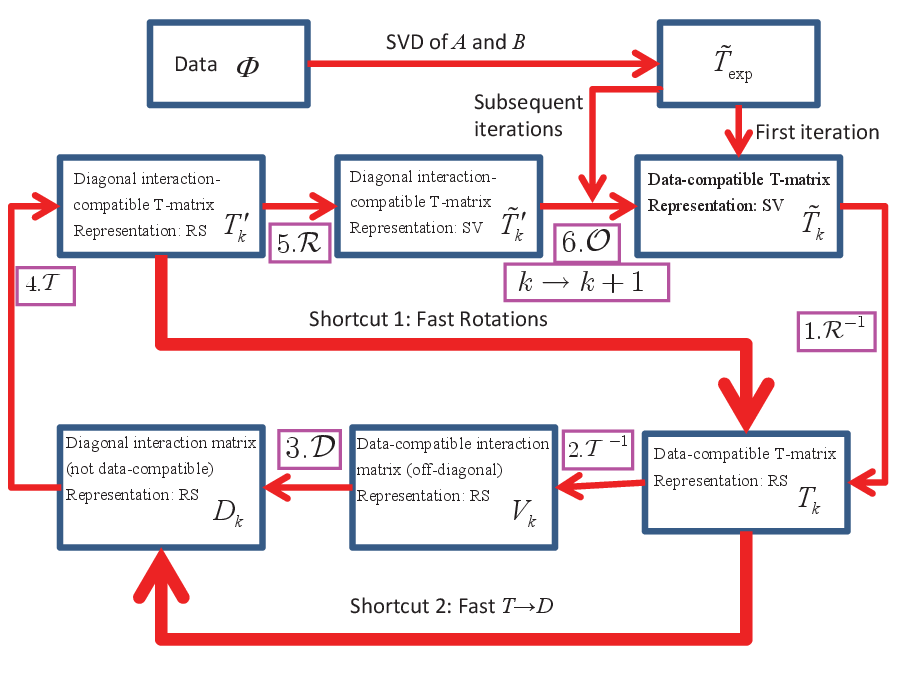}}
\caption{\label{fig:iter} (color online) Basic flowchart of the DCTMC iteration process for the case when the iterations start with an initial guess for $\tilde{T}$. Numbered iteration steps are defined in Sec.~\ref{sec:iter}. Computational shortcuts are described in Sec.~\ref{sec:short}.  Matrix representations are abbreviated by SV (singular-vector) and RS (real-space). Exit condition can be checked at various Steps of the iterations (see~\cite{PRE_2} for more detail).}
\end{figure*}

\begin{enumerate}
 
\item[1:] $T_k= {\cal R}^{-1}[\tilde{T}_k]$\\ 
This transforms the T-matrix from singular-vector to real-space representation. Both $\tilde{T}_k$ and $T_k$ are data-compatible.
  
\item[2:] $V_k = {\cal T}^{-1}[T_k]$\\
This gives $k$-th approximation to the interaction matrix $V$. $V_k$ is data-compatible but not diagonal. 
  
\item[3:] $D_k = {\cal D}[V_k]$\\
Compute the diagonal approximation to $V_k$, denoted here by $D_k$. $D_k$ is diagonal but not data-compatible.
  
\item[4:] $T_k^\prime = {\cal T}[D_k]$\\
Compute the T-matrix that corresponds to the diagonal matrix $D_k$. Unlike $T_k$, $T_k^\prime$ is no longer data-compatible. 
  
\item[5:] $\tilde{T}_k^\prime = {\cal R}[T_k^\prime]$\\
Transform $T_k^\prime$ to singular-vector representation. Here $\tilde{T}_k^\prime$ is still not data-compatible.
  
\item[6:] $\tilde{T}_{k+1} = {\cal O}[\tilde{T}_k^\prime]$\\
Advance the iteration index by one and overwrite the elements of $\tilde{T}_k^{\prime}$ that are known from data with the corresponding elements of $\tilde{T}_{\rm exp}$. This will restore data-compatibility of $\tilde{T}_{k+1}$. Then go to Step 1.

\end{enumerate}

\noindent
The computational complexity of Steps 1,2,4,5 is $O(N_v^3)$. However, the complexity can be dramatically reduced with the use of the computational shortcuts that are described in the next section, with the only exception of Step 4. Therefore, Step 4 is the true computational bottleneck of the method. It's complexity can be reduced by accounting for sparsity of $V$. However, if no {\em a priori} knowledge about sparsity of $V$ is available, then the computational complexity of Step 4 is the limiting factor of DCTMC, at least to the best of our current understanding.

\section{Computational shortcuts}
\label{sec:short}

\subsection{Shortcut 1: Fast pseudo-rotations}
\label{sec:short.rot}

\begin{figure}
\centerline{\includegraphics[width=8.2cm]{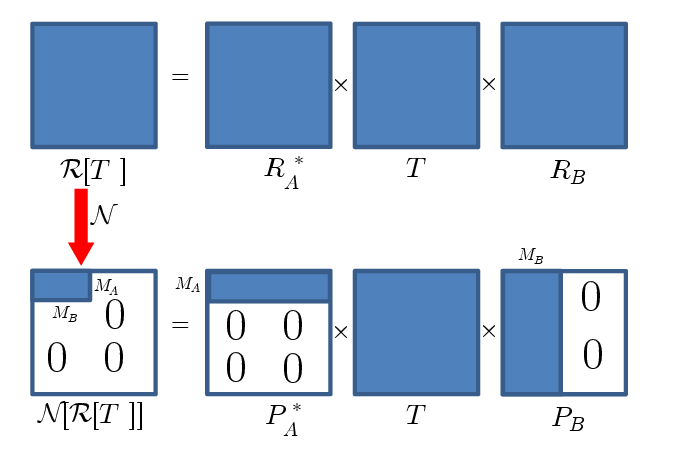}}
\caption{\label{fig:N} (color online) Schematics of computing ${\cal N}\left[ {\cal R}[T] \right]$. Matrices $P_A$ and $P_B$ are obtained from $R_A$ and $R_B$ by setting all columns to zero except for the first $M_A$ and $M_B$ columns, respectively.}
\end{figure}

Consider iteration Steps 5,6,1 written sequentially:
\begin{align*}
& 5: && \tilde{T}_k^\prime = {\cal R}[T_k^\prime]           &      O(N_v^3)   = O(L^9) \ , \\
& 6: && \tilde{T}_{k+1}    = {\cal O}[\tilde{T}_k^\prime]   & \leq O(N_d N_s) = O(L^4) \ , \\
& 1: && T_{k+1}            = {\cal R}^{-1}[\tilde{T}_{k+1}] &      O(N_v^3)   = O(L^9) \ .
\end{align*}
\noindent
To the right, we have indicated the computational complexity of each step and used the previously introduced estimates for $N_d$, $N_s$ and $N_v$ in terms of the grid size $L$. The complexity of Step 6 is equal to, at most, $N_d N_s$. Therefore, the complexity of Steps 5 and 1 is dominating. Now, let us combine the steps by writing
\begin{align}
\label{short_1a}
T_{k+1} &= {\cal R}^{-1} \left[ {\cal O}\left[ {\cal R}[T_k^\prime] \right] \right]  \nonumber \\
        &= {\cal R}^{-1} \left[{\cal R}[T_k^\prime] - {\cal N}\left[ {\cal R}[T_k^\prime] \right] + \tilde{T}_{\rm exp} \right] \ ,
\end{align}
\noindent
where in the second equality we have used the definition of ${\cal O}[\cdot]$ \eqref{O_def}. We now use the linearity and invertibility of ${\cal R}$ to rewrite \eqref{short_1a} as
\begin{align}
\label{short_1c}
T_{k+1} = T_k^\prime + T_{\exp} - {\cal R}^{-1} \left[ {\cal N}\left[ {\cal R}[T_k^\prime] \right] \right] \ .
\end{align}
\noindent
We are therefore left with the task of numerically evaluating an expression of the type ${\cal R}^{-1} [ {\cal N} [ {\cal R}[T] ] ]$, where we have dropped all indexes for simplicity. But this can be accomplished in much less than $O(N_v^3)$ operations due to the sparsity of ${\cal N}[\cdot]$. Indeed, consider first the computation of ${\cal N}[{\cal R}[T]]$. This operation is illustrated in Fig.~\ref{fig:N}. It can be seen that ${\cal N}[{\cal R}[T]] = P_A^* T P_B$, where the matrix $P_A$ is obtained from $R_A$ by overwriting all columns of $R_A$, except for the first $M_A$ columns, with zeros, and $P_B$ is defined analogously. The complexity of computing $P_A^* T P_B$ is $O(\min(M_A,M_B) N_v^2)$, which is less than $N_v^3$ by the factor of at least $O(L)$. Quite analogously, we can show that
\begin{align}
\label{short_1d}
{\cal R}^{-1} [ {\cal N} [ {\cal R}[T_k^\prime] ] ] = P_A\left( P_A^* T P_B \right) P_B^* \ .
\end{align}
\noindent 
It should be kept in mind that $P_A P_A^* \neq I$ and $P_B P_B^* \neq I$ and that premultiplying these matrices is not a computationally efficient approach. Doing so will result in an expression of the type $Q_A^* T Q_B$, where $Q_A = P_A P_A^*$ and $Q_B = P_B P_B^*$ are not sparse. Instead, one should evaluate the right-hand side of \eqref{short_1d} using the operator precedence implied by the parentheses. We conclude that the Steps 5,6,1 of the iterative procedure described above can be combined in the following single computational step: 
\begin{align}
\label{short_1e}
T_{k+1} = T_k^\prime + T_{\exp} - P_A\left( P_A^* T_k^\prime P_B \right) P_B^* \ .
\end{align}
\noindent
In this formula, $T_{\rm exp}$, $P_A$ and $P_B$ are precomputed and stored in memory. We finally note that the sparsity of ${\cal N}$ can be used efficiently even if the region of ''known'' elements of $\tilde{T}$ is not rectangular. Although matrices $P_A$ and $P_B$ can not be easily defined in this case, the use of appropriate masks in computing expressions of the type ${\cal N}[R_A^* T R_B]$ will achieve a similar reduction of computational complexity. 

\subsection{Shortcut 2: Fast $T\rightarrow D$ operation.}
\label{sec:short.T2D}

Here we describe a computational shortcut that can cut the computational time per one iteration by approximately a factor of $\sim 2$. However, we have found empirically~\cite{PRE_2} that this is not the most efficient approach since it does not use one of the main features of DCTMC, that is, accounting for the off-diagonal elements of $V$ at the intermediate stages of the iterations. In~\cite{PRE_2}, it is shown that a more efficient approach is based on utilization of the formula \eqref{D_def} (weighted summation to the diagonal). Here we describe Computational Shortcut 2 for completeness of exposition.

Consider Steps 2 and 3 of the basic iteration cycle:
\begin{align*}
	& 2: && V_k = {\cal T}^{-1}[T_k]  & O(N_v^3) = O(L^9) \ , \\
	& 3: && D_k = {\cal D}[V_k]       & O(N_v)   = O(L^3) \ ,
\end{align*}
\noindent
To the right, we have indicated the computational complexity of each step. The goal of Steps 2 and 3 is to find a diagonal matrix $D$ that in some sense approximates the previously-computed T-matrix. More specifically, we compute $V$ that corresponds to $T$ exactly but is not diagonal in Step 2 and then seek a diagonal approximation to $V$ denoted by $D$. The last operation is governed by the ``force-diagonalization'' operator ${\mathcal D}[\cdot ]$, which in turn depends on the weight function $\rho(\ell)$. As was mentioned above, the simplest choice of this weight function is such that $\rho(\ell_{ki}) = \delta_{ki}$ (we only need to define $\rho(\ell)$ for a set of discrete values of the argument). This choice of the weight function minimizes the entry-wise norm $\Vert V - D \Vert_2$. An alternative approach is to seek a diagonal matrix $D$ that satisfies the equation
\begin{align}
\label{short_2a}
T = D + D{\it \Gamma} T
\end{align}
\noindent
in the minimum $L_2$-norm sense (of course, \eqref{short_2a} can not be satisfied exactly by any diagonal matrix $D$). The above is a classical minimization problem, which has the following analytical solution:
\begin{align}
\label{short_2b}
D_{ij} = \delta_{ij}\frac{T_{ii} + \left[ \left( {\it \Gamma} T \right)^* T \right]_{ii}}
{
1 + \left[\left({\it \Gamma}T\right)^* + \left({\it \Gamma}T \right) + \left({\it \Gamma}T\right)^* \left({\it \Gamma}T \right)
\right]_{ii}
} \ .
\end{align}
\noindent
It may seem that evaluation of \eqref{short_2b} still requires $O(N_v^3)$ operations because it contains the matrix-matrix product ${\it \Gamma}T$. However, this is not so. The matrix ${\it \Lambda} \equiv {\it \Gamma}T$ can be updated iteratively during Computational Shortcut 1 by using \eqref{short_1e} multiplied from the left by ${\it \Gamma}$, viz,
\begin{align}
\label{short_2d}
{\it \Lambda}_{k+1} = {\it \Lambda}_k^\prime + {\it \Lambda}_{\exp} - 
\left(
{\it \Gamma} P_A 
\right)
\left( P_A^* T_k^\prime P_B 
\right) P_B^* \ .
\end{align}
\noindent
Here ${\it \Lambda}_{k+1} = {\it \Gamma}T_{k+1}$, ${\it \Lambda}_k^\prime = {\it \Gamma}T_k^\prime$ and ${\it \Lambda}_{\rm exp} = {\it \Gamma}T_{\rm exp}$. The matrix ${\it \Gamma}P_A$ can be precomputed and has exactly the same sparsity structure as $P_A$ itself, that is, all of its columns except for the first $M_A$ columns are zero. Therefore, computing the last term in \eqref{short_2d} is of the same complexity as Computational Shortcut 1, that is, $O(\min(M_A,M_B)N_v^2)$. There remains the question of how ${\it \Lambda}_k^\prime$ is computed and whether this computation requires an extra matrix-matrix multiplication. The answer is, it can be precomputed at Step 4 of the $k$-th iteration without any additional matrix-matrix multiplications. Indeed, let us utilize the second formula in the definition of ${\cal T}[\cdot]$ \eqref{T_from_V} and write Step 4 of $k$-th iteration as follows:
\begin{align}
\label{short_2e}
T_k^\prime = D_k (I - {\it \Gamma} D_k)^{-1} \ .
\end{align}
\noindent
We then multiply from the left both sides of \eqref{short_2e} by ${\it \Gamma}$ and obtain
\begin{align}
\label{short_2f}
{\it \Lambda}_k^\prime = {\it \Gamma}D_k (I - {\it \Gamma} D_k)^{-1} = (I - {\it \Gamma} D_k)^{-1} - I \ .
\end{align}
\noindent
Now we can compute $T_k^\prime$ and $\Lambda_k^\prime$ without any additional complexity by using the following sub-steps:

\begin{enumerate}

\item[1:] Compute the product ${\it \Delta}_k \equiv {\it \Gamma}D_k$, which is fast because $D_k$ is diagonal. 

\item[2:] Compute the inverse $S_k \equiv (I - {\it \Delta_k})^{-1}$, which has the complexity of $N_v^3$.

\item[3:] Compute ${\it \Lambda}_k^\prime = S_k - I$ [as follows from \eqref{short_2f}].

\item[4:] Compute $T_k^\prime = D_k S_k$ [as follows from \eqref{short_2e}], which is again fast because $D_k$ is diagonal.

\end{enumerate}

Thus, in all the computations outlined above only inversion of $I - {\it \Delta}_k$ has the computational complexity of $N_v^3$. This is, therefore, the true computational bottleneck of the algorithm.

The computational shortcut described here allows one to cut the computational time per one iteration of DCTMC by approximately the factor of $2$. However, we will show in the second part of this paper series~\cite{PRE_2} that a more useful approach is to use an explicit weight function $\rho(\ell)$. This is more in line with the main idea of DCTMC, which is relaxing the requirement that $V$ be a strictly local interaction. Weighted summation to the diagonal is a natural approach to account for the nonlocality (off-diagonality of $V$). Indeed, we will see that, although this approach does not allow one to use the Computational Shortcut 2, it reduces the number of required iterations and is more benefitial in the end.

\subsection{Streamlined iteration cycle}
\label{sec:short.stream}

The computational shortcuts can be integrated into a single streamlined iteration algorithm. Doing so requires careful consideration of the flowchart shown in Fig.~\ref{fig:iter} and of the associated data dependencies. However, the resulting algorithm is relatively simple. For ease of programming, we have broken this algorithm into elementary computational steps. We describe separately the cases when Computational Shortcut 2 is and is not used. In both cases, we start from the initial guess for the T-matrix, $T_1=T_{\rm exp}$. Modification in which the process starts from an initial guess for $V$ is quite obvious and is numerically implemented in~\cite{PRE_2}. \\

\noindent {\bf Initial setup:}

\begin{enumerate}
 
\item[1:] Permanently store in memory the analytically-known matrix ${\it \Gamma}$.

\item[2:] Compute the SVD decomposition \eqref{SVD} of $A$ and $B$. This will yield a set of singular values $\sigma_\mu^A$, $\sigma_\mu^B$ (some of which are identically zero) and singular vectors $\vert f_\mu^A \rangle$, $\vert f_\mu^B \rangle$, $\vert g_\mu^A \rangle$, $\vert g_\mu^B \rangle$.

\item[3:] Use the previous result to construct and permanently store in memory the dense matrices $R_A$ and $R_B$, sparse matrices $P_A$ and $P_B$. If Computational Shortcut 2 is used, compute also $Q_A = {\it \Gamma} P_A$. Note: no additional memory allocation for $P_A$ and $P_B$ is required. 

\item[4:] Compute $\tilde{\it \Phi}_{\mu\nu}$ according to \eqref{Phi_tilde_def} and $\tilde{T}_{\rm exp}$ according to \eqref{T_exp}. Discard the real-space data function, the singular values and singular vectors, and deallocate the associated memory.

\item[5:] Compute and store permanently in memory $T_{\rm exp} = R_A \tilde{T}_{\rm exp} R_B^* = P_A \tilde{T}_{\rm exp} P_B^*$. If computational Shortcut 2 is used, also compute ${\it \Lambda}_{\rm exp} = {\it \Gamma} T_{\rm exp}$.

\item[6:] Initialize iterations by setting $T_1 = T_{\rm exp}$ and ${\it \Lambda_1} = {\it \Lambda}_{\rm exp}$.

\end{enumerate}

\noindent
The computational cost of the initial setup is comparable to that of one iteration or less. Step 1 is negligible. Computation of the SVD of $A$ and $B$ in Step 2 has the cost of $O(N_v(N_s^2 + N_d^2)$, which is also relatively small. The cost of Step 3 is again negligible as it mostly consists of arranging numerical data in the computer memory. The cost of Step 4 is $O(N_s N_d^2 + N_d N_s^2)$. Finally, in Step 5, the complexity of computing $P_A \tilde{T}_{\rm exp} P_B^*$ is $O(N_s N_d^2 + N_d N_s^2)$ or less, depending on the size of nonzero blocks in $P_A$ and $P_B$. The only costly operation is the computation of the matrix-matrix product ${\it \Gamma} T_{\rm exp}$ in Step 5, with the computational complexity of $O(N_v^3)$. But this step is only required if the Computational Shortcut 2 is used.

Exit condition can be defined at different stages of the iterations by using various error measures, as is described in more detail in~\cite{PRE_2}. Here no specific exit conditions are defined.\\

\noindent {\bf Main iteration with the use of Computational Shortcut 2:} Starting from $k=1$, $T_1=T_{\rm exp}$ and ${\it \Lambda_1} = {\it \Lambda}_{\rm exp}$, run the following iterations:

\begin{enumerate}

\item[1:] $\left( D_k \right)_{ij} = \delta_{ij} 
\dfrac{\left( T_k \right)_{ii} + \left( {\it \Lambda}_k^* T_k \right)_{ii}}{1 + \left({\it \Lambda}_k^* + {\it \Lambda}_k + {\it \Lambda}_k^* {\it \Lambda}_k \right)_{ii}}$

\item[2:] ${\it \Delta}_k = {\it \Gamma} D_k$ 

\item[3:] $S_k = \left(I - {\it \Delta}_k \right)^{-1}$

\item[4:] $T_k^\prime = D_k S_k$, \ \ ${\it \Lambda}_k^\prime = S_k - I$

\item[5:] $T_{k+1} = T_k^\prime + T_{\rm exp} - P_A \left( P_A^* T_k^\prime P_B \right) P_B^*$,  \\
   ${\it \Lambda}_{k+1} = {\it \Lambda}_k^\prime + {\it \Lambda}_{\rm exp} - Q_A \left( P_A^* T_k^\prime P_B  \right) P_B^*$

\end{enumerate}

\noindent
Operations whose order of execution is insignificant and which can run independently in parallel threads are shown in the same numbered step. Note that computation of the terms $\left( {\it \Lambda}_k T_k \right)_{ii}$ and $\left( {\it \Lambda}_k^* {\it \Lambda}_k \right)_{ii}$ for $i=1,\ldots,N_v$ has the computational complexity of only $O(N_v^2)$. Therefore, the true bottleneck of each iteration is the operation of matrix inversion $S_k = \left(I - {\it \Delta}_k \right)^{-1}$ whose computational complexity is $O(N_v^3)$.\\

\noindent
{\bf Main iteration without the use of Computational Shortcut 2:} Starting from $k=1$, $T_1=T_{\rm exp}$, run the following iterations:

\begin{enumerate}

\item[1:] $V_k = (I + T_k {\it \Gamma})^{-1} T_k$

\item[2:] $D_k = {\mathcal D}[V_k]$

\item[3:] $T_k^\prime = (I - D_k {\it \Gamma})^{-1} D_k$

\item[4:]  $T_{k+1} = T_k^\prime + T_{\rm exp} - P_A \left( P_A^* T_k^\prime P_B \right) P_B^*$\\

\end{enumerate}

\noindent
The computational cost of this iterative scheme is dominated by the Steps 1 and 3, the computational complexity of each of these steps being  $O(N_v^3)$. 

\section{DCTMC in the linear regime and the questions of convergence and regularization}
\label{sec:lin}

In this section we analyze DCTMC in the linear regime. Most results will be obtained in the case when the weight function in \eqref{D_def} is given by $\rho(\ell_{ki}) = \delta_{ki}$. However, generalizations for a more general (and a more practical) operator ${\mathcal D}[\cdot]$ will be mentioned briefly.

Consider the iteration cycle of Sec.~\ref{sec:short.stream} in the limit ${\it \Gamma} \rightarrow 0$. Omitting intermediate steps, we find that each iteration is reduced to the following two operations:

\begin{enumerate}

\item[1:] $D_k = {\cal D}[T_k]$

\item[2:] $T_{k+1} = D_k + T_{\rm exp} - \left( P_A  P_A^* \right) D_k \left( P_B P_B^* \right)$

\end{enumerate}

\noindent
where ${\cal D}[\cdot]$ is defined in \eqref{D_def}. These two steps can be combined in the following simple iteration:
\begin{align}
\label{D_iter_lin}
D_{k+1} = D_k + D_{\rm exp} - {\cal D}\left[ \left( P_A  P_A^* \right) D_k \left( P_B P_B^* \right) \right] \ ,
\end{align}
\noindent
where $D_{\rm exp} = {\cal D}[T_{\rm exp}]$. Iteration \eqref{D_iter_lin} can be obtained simply by applying the operator ${\cal D}[\cdot]$ to the equation in Step 2 above. Let us now convert \eqref{D_iter_lin} to an equation with respect to the vector $\vert \upsilon_k \rangle$ that contains the diagonal elements of $D_k$. From the linearity of \eqref{D_iter_lin} we immediately find
\begin{align}
\label{v_iter_lin}
\vert \upsilon_{k+1} \rangle = \vert \upsilon_{\rm exp} \rangle + (I - W) \vert \upsilon_k \rangle \ ,
\end{align}
\noindent
where $\vert \upsilon_{\rm exp} \rangle$ is the vector of diagonal elements of $D_{\rm exp}$.

We now specialize to the case when the weight function in the definition \eqref{D_def} of the operator ${\mathcal D}[\cdot]$ is given by $\rho(\ell_{ki}) = \delta_{ki}$. In this case, the matrix $W$ has the elements
\begin{align}
\label{M_def}
W_{ij} = \left( P_A  P_A^* \right)_{ij} \left( P_B P_B^* \right)_{ji} \ .
\end{align}
\noindent
It is easy to see that \eqref{v_iter_lin} is Richardson first-order iteration with the fixed point $\vert \upsilon_\infty \rangle = W^{-1} \vert \upsilon_{\rm exp} \rangle$. Therefore, DCTMC in the linear regime simply provides an iterative way of solving the equation 
\begin{align}
\label{M_upsilon_exp}
W \vert \upsilon \rangle= \vert \upsilon_{\rm exp} \rangle \ . 
\end{align}
\noindent
This equation can be derived independently from DCTMC and in a more straightforward manner starting from the linearized equation \eqref{K_phi_lin}. This derivation is shown in Appendix~\ref{app:M} and it takes advantage of the algebraic properties of $K$ (see Remark 3). It is important to realize that, although \eqref{M_upsilon_exp} can be obtained from \eqref{K_phi_lin} by a series of linear transformations, the two equations are not equivalent in the following sense: if $K$ is not invertible, then the pseudoinverse solutions of the two equations can be different. However, if $K$ is invertible, then the two equations have the same unique solution. 

Of course, iteration \eqref{v_iter_lin} is only a particular numerical method of solving \eqref{M_upsilon_exp} and not the most efficient one: conjugate-gradient descent is expected to provide better computational performance. However, consideration of DCTMC in the linear regime is not a vane or trivial exercise but is useful in several respects. First, it gives us an insight into the convergence properties of DCTMC. Second, it gives us an idea of how DCTMC iterations can be regularized. Convergence and regularization will be discussed in the remainder of this section.

It is obvious that the iterations converge to the fixed point provided that $\vert 1 - w_n \vert < 1$ for all $n$, where $w_n$ are the eigenvalues of $W$. Since $W$ is Hermitian, all its eigenvalues are real and therefore the convergence condition reads $0 < w_n < 2$. Under the same condition the inverse $W^{-1}$ exists. We will now prove that 
\begin{align}
\label{lambda_ineq}
0 \le w_n \leq 1 \ .
\end{align}
\noindent
The fact that $W$ is non-negative definite is obvious from \eqref{M_def}. We will however make an additional step and recall that the columns of $P_A$ are the singular vectors $\vert g_\mu^A \rangle$ for $\mu=1,\ldots,M_A$ and zeros otherwise while the columns of $P_B$ are the singular vectors $\vert f_\mu^B \rangle$ for $\mu=1,\ldots,M_B$ and zeros otherwise. We then obtain in a straightforward manner:
\begin{align}
\label{M_f_g}
W_{ij} = \sum_{\mu=1}^{M_A}\sum_{\nu=1}^{M_B}  \langle i \vert g_\mu^A \rangle \langle g_\mu^A \vert j \rangle \langle j \vert f_\nu^B \rangle \langle f_\nu^B \vert i \rangle \ .
\end{align}
\noindent
Let $\vert x \rangle$ be an arbitrary nonzero vector of length $N_v$ and $X$ be an $N_v \times N_v$ matrix with the elements of $\vert x \rangle$ on the diagonal and zeros elsewhere. Then
\begin{align}
\label{x_M_x}
\langle x \vert W  \vert x \rangle = \sum_{\mu=1}^{M_A} \sum_{\nu=1}^{M_B} \left\vert  \langle g_\mu^A \vert X \vert f_\nu^B \rangle \right\vert^2 \ge 0 \ .
\end{align}
\noindent
We therefore have proved that $w_n \ge 0$. Next, we use the orthonormality of each set of singular vectors to write the following identities
\begin{align}
\label{x_x}
\langle x \vert x \rangle = \sum_{i=1}^{N_v} \left\langle i \vert X^* X \vert i \right\rangle = \sum_{\mu=1}^{N_v} \sum_{\nu=1}^{N_v}  \left\vert \langle g_\mu^A \vert X \vert f_\nu^B \rangle  \right\vert^2 \ .
\end{align}
\noindent
Since $M_a,M_b \leq N_v$, $\langle x \vert W \vert x \rangle \leq \langle x \vert x \rangle$ and we have proved \eqref{lambda_ineq}. The equality $\langle x\vert W \vert x \rangle = \langle x \vert x \rangle$ holds only in the case $M_A = M_B = N_v$, in which case $W=I$ and the iteration \eqref{v_iter_lin} trivially converges to its fixed point right upon making the initial guess. In this unrealistic case, all elements of the T-matrix are determined from the data and no iterations are needed.

We thus conclude that convergence can be slow in the case $W$ has a small (or zero) eigenvalue. We can define the characteristic {\em overlap} of singular vectors related to detectors and sources as
\begin{align}
\xi = \inf_{X\neq 0}\left\{ \dfrac{\sum_{\mu=1}^{M_A} \sum_{\nu=1}^{M_B} \left\vert  \langle g_\mu^A \vert X \vert f_\nu^B \rangle  \right\vert^2}{\sum_{\mu=1}^{N_v} \sum_{\nu=1}^{N_v} \left\vert  \langle g_\mu^A \vert X \vert f_\nu^B \rangle  \right\vert^2} \right\}  \ .
\end{align}
\noindent
The iterations \eqref{v_iter_lin}  converge at least as fast as the power series $\sum_n (1 - \xi)^n$. If $\xi$ is close to zero, the convergence can be slow. This observation gives us an idea of how the iterations can be regularized. This can be accomplished by replacing $W$ by  $W + \lambda^2 I$, where $\lambda$ is a regularization parameter. As is shown in Appendix~\ref{app:M}, this procedure is equivalent to Tikhonov regularization of equation \eqref{Q_tau}, which can be derived from \eqref{K_phi_lin} by several linear operations. However, the substitution $W \rightarrow W + \lambda^2 I$ is not equivalent to Tikhonov regularization of \eqref{K_phi_lin}. 

We can now introduce regularization of the general iterative algorithm of Sec.~\ref{sec:short.stream}, which is applicable to the nonlinear case as well. Namely, for any matrix $X$, we replace the linear transformation $(P_A P_A^*) X (P_B P_B^*)$ by $(P_A P_A^*) X (P_B P_B^*) + \lambda^2 {\cal D}[X]$. This entails the following modification to Step 5 of the streamlined algorithm with Shortcut 2 or Step 4 of the algorithm without Shortcut 2:  
\begin{align*}
& T_{k+1} = T_k^\prime - \lambda^2 {\cal D}\left[ T_k^\prime \right] + T_{\rm exp} - P_A \left( P_A^* T_k^\prime P_B \right) P_B^* \ , \\ 
& {\it \Lambda}_{k+1} = {\it \Lambda}_k^\prime  - \lambda^2 {\cal D}\left[ {\it \Lambda}_k^\prime \right] + {\it \Lambda}_{\rm exp} - Q_A \left( P_A^* T_k^\prime P_B \right) P_B^* \ .
\end{align*}

Finally, we mention briefly which modifications can be expected if we use a more general weight function $\rho(\ell_{ki})$ in the definition \eqref{D_def} of the operator ${\mathcal D}$. The first obvious result is that the matrix $W$ is modified in this case so that its elements are given by
\begin{align}
\label{W_mod}
W_{ij} &= \sum_k (P_A P_A^*)_{ij} (P_B P_B^*)_{jk} \rho(\ell_{ki}) \nonumber \\
       &=  (P_A P_A^*)_{ij} (P_B P_B^* H)_{ji} \ ,
\end{align}
\noindent 
where $H_{ij} = \rho(\ell_{ij})$. This matrix is no longer non-negative definite. Moreover, its eigenvalues can be complex and we can not prove easily that they are constrained to the disk $\vert 1 - w_n \vert < 1$ (in fact, this may be not so). However, numerical evidence shows that using weight functions $\rho(\ell)$ that are non-zero for finite $\ell$ improves the convergence rate of the DCTMC iterations~\cite{PRE_2}. We view this as an indication that the eigenvalues $w_n$ are pushed away from the origin in the complex plane while staying within the disk $\vert 1 - w_n \vert < 1$. The free term of the equation \eqref{M_upsilon_exp} is also affected by the choice of the weight function $\rho(\ell)$. Indeed, we can write $\langle i \vert \upsilon_{\rm exp} = (T_{\rm exp} H)_{ii}$. The resultant equation $W\vert \upsilon \rangle = \vert \upsilon_{\rm exp} \rangle$ is still derivable from \eqref{K_phi_lin}. In Appendix\ref{app:M}, the derivation is shown for the simple case $\rho(\ell_{ki}) = \delta_{ki}$ or $H=I$, but the more general case can be easily considered and the resultant transformation between $K$ and $W$ is given in the end of the Appendix. 

\section{Discussion}
\label{sec:disc}

This paper describes a novel method for solving nonlinear inverse scattering problems (ISPs). The method is based on iterative completion of the unknown entries of the T-matrix and we refer to it as to the data-compatible T-matrix completion (DCTMC) method. It should be emphasized that the constraint that we apply to the T-matrix (namely, that it corresponds to a nearly diagonal interaction matrix $V$) is not the same as in the conventional formulation of the matrix completion problem. The method developed in this paper is well suited for overdetermined ISPs in which the number of volume voxels is not too large (e.g., $\lesssim 10^4$) or the target is sparse. The size of the data set in not a limiting factor for this method, unlike in many traditional approaches to the same problem.

In the case of ill-posed ISPs, regularization plays the key role. One should not expect to recover a reasonable image without some form of regularization. DCTMC allows for two types of regularization: (i) by imposing physical constraints and (ii) by regularizing the matrix $W_{ij} = (P_A P_A^*)_{ij} (P_B P_B^*)_{ji}$. In the linear regime, the approach (ii) corresponds to Tikhonov regularization of the linearized equation ${\it \Theta}U K \vert \upsilon \rangle = {\it \Theta}U \vert \phi \rangle$, which is obtained from Eq.~\eqref{K_phi_lin} by multiplying the latter by ${\it \Theta}U$ from the left; the unitary matrix $U$ and the matrix of diagonal scaling ${\it \Theta}$ are defined in Appendix~\ref{app:M}. In the nonlinear regime, the approach (ii) is somewhat {\em ad hoc} and its applicability requires additional research.

A potential advantage of DCTMC is its computational efficiency in solving strongly overdetermined inverse problems with large data sets. This advance is obtained by exploiting the algebraic structure of the ISP rather than stating it in the conventional generic form $F[\upsilon] = 0$ (or $K\upsilon=0$ in the linear approximation) where $F[\cdot]$ is a general nonlinear functional, $K$ is its linear approximation and $\upsilon$ is the vector consisting of the diagonal elements of $V$. In our previous work, we have shown that strongly overdetermined data sets are required to obtain the optimal image resolution~\cite{markel_02_2,markel_04_4}. Specifically, the fundamental limit of {\em lateral} resolution of diffuse optical tomography (DOT) in the slab geometry is equal to the step $h$ of the mesh of sources and detectors on each face of the slab, provided that the imaging windows are significantly larger than the field of view of the reconstruction. Thus, to reconstruct a lateral central cross section of the slab on a $100\times 100$ mesh of step $h$, one needs about $300^2$ sources on one side and the same number of detectors on the other side of the slab. This translates into $\sim 10^{10}$ data points. This result is not specific to DOT but also holds for diffraction tomography that we consider numerically in~\cite{PRE_2}. However, the fundamental resolution limit is not always achievable. If the inverse problem is ill-posed or noise is present in the data, the theoretical limit of resolution can not be achieved and smaller data sets can suffice to obtain the optimal image quality. Therefore, the optimal size of the data set is determined by a complex interplay between the ill-posedness of the inverse problem and the statistical properties of the noise. Experimental DOT reconstruction with large data sets was demonstrated in~\cite{wang_05_1}, experimental determination of the optimal size of the data set was  in~\cite{wang_05_1,konecky_08_1} and further insights on selection of useful data points in the presence of strong nonlinearity were provided in~\cite{ban_13_1}. However, all references just quoted used linearized image reconstruction. DCTMC allows one to overcome this limitation and obtain nonlinear reconstructions with very large data sets.

Although the main goal of DCTMC is to solve nonlinear problems, the methods developed above can be useful for solving linearized problems with large data sets as well. This development is conceptually similar to the image reconstruction methods of~\cite{schotland_97_1,schotland_01_1,markel_03_1,markel_04_4} that were developed for solving linear ISPs. The similarity lies in exploring the algebraic structure of the matrix $K$, which is obtained as a product of two unperturbed Green's functions. In~\cite{schotland_97_1,schotland_01_1,markel_03_1,markel_04_4} the special structure of $K$ that follows from the translational invariance of an infinite homogeneous slab was exploited. In this work, we do not use or assume translational invariance of the medium and do not work in the infinite slab geometry. Instead, we obtain an expression of the form $AVB$, where $V$ is the unknown potential (in the linear regime, $V=T$). This replaces the traditional approach in which one writes $AVB = K\upsilon$, where $\upsilon$ is the vector of diagonal elements of $V$. 

In the full nonlinear regime, DCTMC also bears some similarity to the methods of~\cite{chaumet_04_1,mudry_12_1,jakobsen_15_1} where the notion of the fundamental unknown is also expanded to include the internal fields or the complete Green's function $G$ or the T-matrix $T$. For example, in the work of  Chaumet {\em et al.}, the {\em dipole moments} ${\bf d}_n$ of voxels are iteratively updated. In our terminology, $d_{n\alpha} = (Tq)_{n\alpha}$, where $\alpha=x,y,z$ labels Cartesian components of three-dimensional vectors and $q$ is the incident field (see Sec.~\ref{sec:gen_ISP}). The intermediate dipole results are directly updated by a step in the Polak-Ribiere conjugate gradient direction. Then, once an acceptable result has been obtained, the relationship between the dipoles and the voxel polarizabilities is used to obtain the final reconstruction. DCTMC is different in several respects. It searches in a different direction, which is determined at each iteration by the {\it experimental T-matrix}. Also, the voxel polarizabilities are updated at {\it each} iteration (here we imply a unique one-to-one correspondence between the voxel polarizabilities and the T-matrix). Therefore, our treatment of the unknowns is similar to that in Refs.~\cite{chaumet_04_1,mudry_12_1,jakobsen_15_1} but the method for updating the unknown is different. \\

\section*{Acknowledgments}

This work has been carried out thanks to the support of the A*MIDEX project (No. ANR-11-IDEX-0001-02) funded by the ``Investissements d'Avenir'' French Government program, managed by the French National Research Agency (ANR) and was also supported by the US National Science Foundation under Grant DMS 1115616 and by the US National Institutes of health under Grant P41 RR002305. The authors are grateful to J.C.Schotland, A.Yodh and Anne Sentenac for very useful discussions.

\bibliographystyle{apsrev}
\bibliography{abbrev,book,master,local}

\begin{thebibliography}{39}
\bibitem[{fn1()}]{fn1}
\bibinfo{note}{If the source is not of unit strength, measurements should be
  divided by the source amplitude.}
\bibitem[{fn2()}]{fn2}
\bibinfo{note}{There is no real distinction between the two cases if the data
  are sampled.}
\bibitem[{fn3()}]{fn3}
\bibinfo{note}{This possibility is closely related to the existence of
  non-radiating sources. Indeed, to force the rank of $B$ to be less than
  $N_s$, we need to arrange the sources in space and to assign them relative
  amplitudes and phases so that they produce almost no field in $\Omega$.}
\end{thebibliography}

\appendix
\section{Linearizing approximations}
\label{app:lin}

In this Appendix we state the linearizing approximations without derivation or analysis as this is outside of the scope of this paper. We only note that first Born approximation is obtained by retaining the first-order term in the power-series expansion of the complete Green's function $G$; first Rytov approximation is obtained by retaining the first-order term in the cumulant expansion of $\log(G)$ and the mean-field approximation is obtained by using the first-order approximant in the continued-fraction expansion of $G$. More details are given in~\cite{markel_04_4}. Only first Born approximation can be stated in abstract form while the other two approximations involve entry-wise expressions. Correspondingly, the accuracy of these two approximations depends on the matrix representation while the accuracy of first Born approximation is representation-independent.

{\bf (i) First Born approximation.} The simplest approach to linearization is the first Born approximation, according to which
\begin{align}
A (I - V{\it \Gamma})^{-1} V B \approx A V B \ . \nonumber
\end{align}
\noindent
Obviously, the first Born approximation is valid if $\Vert V {\it \Gamma} \Vert_2 \ll 1$. By substituting the above approximation into the left-hand side of \eqref{Phi_V_matrix}, we obtain \eqref{Psi_V_matrix_lin} in which ${\it \Psi} = {\it \Phi}$. Therefore, the data transformation in the case of first Born approximation is trivial.

{\bf (ii) First Rytov approximation.} The first Rytov approximation can be stated as
\begin{align}
\left[ A (I - V {\it \Gamma})^{-1} V B \right]_{ij} \approx C_{ij} 
\left\{
\exp\left[ \dfrac{(AVB)_{ij}}{C_{ij}}  \right] - 1
\right\} \ , \nonumber
\end{align}
\noindent
where $C_{ij} = G_0({\bf r}_i, {\bf r}_j)$ and ${\bf r}_i \in \Sigma_d$, ${\bf r}_j \in \Sigma_s$. Thus we have encountered yet another restriction of $G_0({\bf r},{\bf r}^\prime)$. Obviously, with this restriction used, $G_0({\bf r}, {\bf r}^\prime)$ is the {\em direct} (unscattered) field that would have been produced by a source located at ${\bf r}$ and measured by a detector at ${\bf r}^\prime$ in the case $V=0$. If we substitute the above approximation into the left-hand side of \eqref{Phi_V_matrix} and introduce the data transformation 
\begin{align}
{\it \Psi}_{ij} = C_{ij} \log \left( 1 + {\it \Phi_{ij}}/C_{ij} \right) \ , \nonumber
\end{align}
\noindent
we would arrive again at \eqref{Psi_V_matrix_lin}. Therefore, the equation above defines the data transformation of first Rytov approximation.

{\bf (iii) Mean-field approximation.} The mean-field approximation is 
\begin{align}
\left[ A (I - V {\it \Gamma})^{-1} V B \right]_{ij} \approx 
\dfrac{(AVB)_{ij}}{1 - (AVB)_{ij}/C_{ij}} \ . \nonumber
\end{align}
\noindent
The data transformation of the mean-field approximation has the form of element-wise harmonic average and reads
\begin{align}
{\it \Psi}_{ij} = \dfrac{1}{1/{\it \Phi}_{ij} + 1/C_{ij} } \ . \nonumber
\end{align}

\section{Derivation of the equation $W \vert \upsilon \rangle = \vert \upsilon_{\rm exp} \rangle$ [Eq.~\eqref{M_upsilon_exp}] from Eq.~\eqref{K_phi_lin}.}
\label{app:M}

In this Appendix, we derive Eq.~\eqref{M_upsilon_exp}] from Eq.~\eqref{K_phi_lin} for the special case of the weight function $\rho(\ell_{ki}) = \delta_{ki}$. The more general case can be considered without difficulty, and the relevant result is adduced in the end of this appendix.

Consider the linearized equation \eqref{K_phi_lin} in the first Born approximation, that is, with the trivial data transformation 
${\it \Psi} = {\it \Phi}$ or equivalently $\vert \psi \rangle = \vert \phi \rangle$. We recall that $K_{(mn),j} = A_{mj} B_{jn}$ 
and use \eqref{SVD} for $A_{mj}$ and $B_{jn}$ to obtain the following result for the elements of $K$:
\begin{align}
K_{(mn),j} = \sum_{\mu=1}^{N_d}\sum_{\nu=1}^{N_s} \sigma_\mu^A \sigma_\nu^B \langle m \vert f_\mu^A \rangle \langle g_\mu^A \vert j \rangle \langle j \vert f_\nu^B \rangle \langle g_\nu^B \vert n \rangle \ . \nonumber
\end{align}
\noindent
Now define the unitary matrix $U$ with the elements
\begin{align}
\label{App_C_2}
U_{(\mu\nu),(mn)} &= \langle f_\mu^A \vert m \rangle \langle n \vert g_\nu^B \rangle \ , \\ 
& 1\leq \mu,m \leq N_d \ , \ 1 \leq \nu,n \leq N_s \nonumber
\end{align}
\noindent
and multiply \eqref{K_phi_lin} by $U$ from the left:
\begin{align}
\label{App_C_3}
(UK) \vert \upsilon \rangle = U\vert \phi \rangle \ .
\end{align}
\noindent
We emphasize that multiplication of any linear equation by a unitary matrix does not change its Tikhonov-regularized pseudoinverse solution. This follows immediately from $(UK)^*(UK) = K^* K$ and $(UK)^*U = K^*$. Now we use the equalities
\begin{align}
(UK)_{(\mu\nu),j} = \sigma_\mu^A \sigma_\nu^B \langle g_\mu^A \vert j \rangle \langle j \vert f_\nu^B \rangle \ , \
\langle (\mu\nu) \vert U \vert \phi \rangle = \tilde{\it \Phi}_{\mu\nu} \nonumber
\end{align}
\noindent
to re-write \eqref{App_C_3} as
\begin{align}
\sigma_\mu^A \sigma_\nu^B \sum_{j=1}^{N_v} \langle g_\mu^A \vert j \rangle \langle j \vert f_\nu^B \rangle \langle j \vert \upsilon \rangle = \tilde{\it \Phi}_{\mu\nu} \ ,\nonumber  \\ 
1\leq \mu \leq N_d \ , \  1 \leq \nu \leq N_s \ . \nonumber
\end{align}
\noindent
Next we observe that the above set may contain some equations in which all coefficients are zero or very small. These equations can be safely discarded and this operation still does not affect the pseudo-inverse. Therefore we obtain
\begin{align}
\label{App_C_6}
\sigma_\mu^A \sigma_\nu^B \sum_{j=1}^{N_v} \langle g_\mu^A \vert j \rangle \langle j \vert f_\nu^B \rangle \langle j \vert \upsilon \rangle = \tilde{\it \Phi}_{\mu\nu} \ , \\ 
1\leq \mu \leq M_A \ , \ \  1 \leq \nu \leq M_B  \ , \nonumber
\end{align}
\noindent
where $M_A$ and $M_B$ are the dimensions of the shaded rectangle in Fig.~\ref{fig:T}, which is the region where the inequality $\sigma_\mu^A \sigma_\nu^B > \epsilon^2$ holds and $\epsilon$ is the small positive constant introduced in \eqref{T_inv}. Note that \eqref{App_C_6} is in all respects equivalent to \eqref{K_phi_lin}.

At this point however, we make a transformation that will, in fact, change the equation. Namely, we divide \eqref{App_C_6} by the factor $\sigma_\mu^A \sigma_\nu^B$, which is larger than $\epsilon^2$ for all equations included in \eqref{App_C_6}. In computational linear algebra, this operation is known as preconditioning by diagonal scaling. We thus obtain
\begin{align}
\label{App_C_7}
\sum_{j=1}^{N_v} \langle g_\mu^A \vert j \rangle \langle j \vert f_\nu^B \rangle \langle j \vert \upsilon \rangle = \left( \tilde{T}_{\rm exp} \right)_{\mu\nu} \ , \\ 
1\leq \mu \leq M_A \ , \ \  1 \leq \nu \leq M_B \ , \nonumber
\end{align}
\noindent
where we have used the definition \eqref{T_exp} of $\tilde{T}_{\rm exp}$. The diagonal scaling that was applied to obtain \eqref{App_C_7} is invertible. Therefore, if \eqref{K_phi_lin} has a solution in the ordinary sense, then \eqref{App_C_7} has the same solution. However, if \eqref{K_phi_lin} is not invertible, then the two equations have different pseudoinverse solutions that can not be related to each other by a simple transformation. In this sense, the two equations \eqref{K_phi_lin} and \eqref{App_C_7} are no longer equivalent.

To obtain \eqref{M_upsilon_exp}, we observe that \eqref{App_C_7} can be written as 
\begin{align}
\label{Q_tau}
Q \vert \upsilon \rangle = \vert \tau \rangle \ ,
\end{align}
\noindent
where $Q_{(\mu\nu),j} = \langle g_\mu^A \vert j \rangle \langle j \vert f_\nu^B \rangle$ and $\langle (\mu\nu) \vert \tau \rangle = \left( \tilde{T}_{\rm exp} \right)_{\mu\nu}$. We then multiply \eqref{App_C_6} by $Q^*$ from the left and obtain
\begin{align}
\left( Q^* Q \right)_{ij} = \sum_{\mu=1}^{M_A} \sum_{\nu=1}^{M_B} \langle f_\nu^B \vert i \rangle \langle i \vert g_\mu^A \rangle \langle g_\mu^A \vert j \rangle \langle j \vert f_\nu^B \rangle = W_{ij} \ , \nonumber
\end{align}
\noindent
where $W = Q^*Q$ is the same matrix as in \eqref{M_def} (compare the above equation to \eqref{M_f_g}). In a similar manner, we obtain 
\begin{align}
\langle i \vert Q^* \vert \tau \rangle = \sum_{\mu_1}^{M_A} \sum_{\nu=1}^{M_B} Q^*_{i,(\mu\nu)} \left(\tilde{T}_{\rm exp}\right)_{\mu\nu} \nonumber \\
= (P_A\tilde{T}_{\rm \exp} P_B^* )_{ii} = (T_{\rm exp})_{ii} =  \langle i \vert \upsilon_{\rm exp} \rangle \ . \nonumber
\end{align}
\noindent
Therefore, $Q^*\vert \tau \rangle = \vert v_{\rm exp} \rangle$. We thus conclude that \eqref{M_upsilon_exp} is obtained from \eqref{Q_tau} by multiplying both sides with $Q^*$. Moreover, the substitution $W \rightarrow W + \lambda^2 I$ is equivalent to Tikhonov-regularization of \eqref{Q_tau}.

Finally, we can state the formal relation between $K$ and $W$ in the following form:
\begin{align}
W = ({\it \Theta} UK)^*({\it \Theta} UK) = K^* U^{-1} {\it \Theta}^2 U K \ , \nonumber
\end{align}
\noindent
where $U$ is the unitary matrix defined in \eqref{App_C_2} and ${\it \Theta}$ is the diagonal conditioning matrix containing the quantities $1/(\sigma_\mu^A \sigma_\nu^B)$ for $1 \leq \mu \leq M_A$ and $1 \leq \nu \leq M_B$ and zeros otherwise.

Above consideration applied to the case $\rho(\ell_{ki}) = \delta_{ki}$ or, equivalently, $H = I$, where $H_{ki} = \rho(\ell_{ki})$. If $H\neq I$, the matrix $W$ is given by
\begin{align}
W = ({\it \Theta} UK)^* F ({\it \Theta} UK) \ . \nonumber
\end{align}
\noindent
The matrix $F$ appears in this expression as an additional filter. The free term of the equation is modified so that $\langle i \vert \upsilon_{\rm exp} \rangle = (T_{\rm exp} F)_{ii}$.

\section{Definitions and properties of several functionals used in this paper}
\label{app:oper}

A summary of definitions and mathematical properties of the several functionals used in this paper is given in Table~\ref{tab:1}. In this table, ${\cal F}$ refers to any of the functionals ${\cal T}$, ${\cal R}$, ${\cal D}$, ${\cal M}$ and ${\cal N}$ and ${\cal O}$.

\begin{table*}
\begin{tabular}{|c|c|l|c|l|}
	\hline
	${\cal F}$ &  Entry-wise? & $Y={\cal F}[X]$ & Invertible? & $X={\cal F}^{-1}[Y]$ \\
	\hline
	${\cal T}$ & No & $Y = (I-X{\it \Gamma})^{-1}X$ & Sometimes & $X = (I + Y{\it \Gamma})^{-1}Y$ \\
	\hline
	${\cal R}$ & No  & $Y = R_A^*X R_B$ & Yes & $X = R_AYR_B^*$ \\
	\hline
	${\cal D}$ & Yes & $Y_{ij} = \delta_{ij} \sum_k X_{ik} \rho(\ell_{ki})  $ & No & N/A \\ 
	\hline
	${\cal M}$ & Yes & $\tilde{Y}_{\mu\nu} = 
	\left\{
	\begin{array}{ll}
	0                  \ , & \sigma_\mu^A \sigma_\nu^B >\epsilon^2 \\
	\tilde{X}_{\mu\nu} \ , & {\rm otherwise}
	\end{array} \right. $
	& No & N/A \\ 
	\hline
	${\cal N}$ & Yes & $\tilde{Y}_{\mu\nu} = 
	\left\{
	\begin{array}{ll}
	\tilde{X}_{\mu\nu} \ , & \sigma_\mu^A \sigma_\nu^B >\epsilon^2 \\
	0                  \ , & {\rm otherwise}
	\end{array} \right. $
	& No & N/A \\ 
	\hline
    &	&	&	&	\\
	${\cal O}$ & Yes & $\tilde{Y} = {\cal M}[\tilde{X}] + \tilde{T}_{\rm exp}$ & No & N/A \\ 
	           &     & \hspace*{3.0mm} = $\tilde{X} - {\cal N}[\tilde{X}] + \tilde{T}_{\rm exp}$ & & \\
	\hline
\end{tabular} \\
\vspace*{2mm}
\caption{\label{tab:1} Definitions and properties of the various functional used in this paper. The weight function $\rho(\ell)$ must be defined separately and is expected to go to zero for large values of $\ell$; $\ell_{ki}$ is the physical distance between voxels $k$ and $i$.} 
\end{table*}

\end{document}